\documentclass [11pt, letterpaper]{article}
\pdfoutput=1 

\usepackage{fullpage}
\usepackage{amsmath}
\usepackage{amssymb}
\usepackage{graphicx}
\usepackage{mathrsfs}
\usepackage{wrapfig}
\usepackage{boxedminipage}
\usepackage{epsfig}
\usepackage{setspace}
\usepackage{subfigure}
\usepackage{float}
\usepackage{hyperref}
\usepackage{enumerate}
\usepackage{cite}
\usepackage[font=footnotesize,labelfont=rm]{caption}

\begin{document}
\begin{flushright}
{\tt arXiv:1612
}
\end{flushright}

\bigskip
\bigskip
\bigskip
\bigskip

\bigskip
\bigskip
\bigskip
\bigskip

\begin{center}
{\Large
{\bf Benchmarking Black Hole Heat Engines
}
}
\end{center}

\bigskip
\begin{center}
{\bf  Avik Chakraborty and
Clifford V. Johnson}

\end{center}

\bigskip
\bigskip

\begin{center}
  \centerline{\it Department of Physics and Astronomy }
\centerline{\it University of
Southern California}
\centerline{\it Los Angeles, CA 90089-0484, U.S.A.}
\end{center}

\bigskip
\centerline{\small \tt avikchak, johnson1,  [at] usc.edu}

\bigskip
\bigskip
\bigskip

\begin{abstract} 
\noindent  
We present the results of initiating a benchmarking scheme that allows for  cross--comparison of the efficiencies of black holes used as working substances in heat engines.  We use a circular cycle  in the $p{-}V$ plane as the benchmark engine. We test it on Einstein--Maxwell, Gauss--Bonnet, and Born--Infeld black holes. Also, we derive  a new  and surprising exact result for the efficiency of a special ``ideal gas" system to which all the black holes asymptote. 
\end{abstract}
\newpage \baselineskip=18pt \setcounter{footnote}{0}

\section{Introduction}
\label{Intro} 

The classic black hole thermodynamics \cite{Bekenstein:1973ur,Bekenstein:1974ax,Hawking:1974sw,Hawking:1976de}  relates the mass $M$, surface gravity $\kappa$, and outer horizon area $A$ of a black hole solution to the  energy, temperature, and entropy  ($U$, $T$, and  $S$, resp.) according to\footnote{Here we are using geometrical units where $G,c,\hbar, k_{\rm B}$ have been set to unity.}: 
\begin{eqnarray} \label{oldTD}
M=U \ , \ T=\frac{\kappa}{2 \pi} \ ,\ S=\frac{A}{4} \ .
\end{eqnarray}
 The formalism has been extended\footnote{For a selection of references, see refs. \cite{Caldarelli:1999xj,Wang:2006eb,Sekiwa:2006qj,LarranagaRubio:2007ut,Kastor:2009wy,Dolan:2010ha,Cvetic:2010jb,Dolan:2011jm,Dolan:2011xt}, including the reviews in refs. \cite{Dolan:2012jh,Altamirano:2014tva,Kubiznak:2016qmn}. See also the early work in refs. \cite{Henneaux:1984ji,Teitelboim:1985dp,Henneaux:1989zc}.} by allowing the cosmological constant of the theory to be dynamical, supplying a pressure {\it via}  $ p=-\Lambda/8\pi$, along with its conjugate volume $V$. Now, the black hole mass is  related to the enthalpy $H$ of the system instead of the energy $U$\cite{Kastor:2009wy}: $M=H\equiv U+pV$. The First Law now reads as:
\begin{eqnarray} \label{newTD}
dM=TdS+Vdp+\Phi_{i} dq_{i}+\Omega_{i} dJ_{i}\ .
\end{eqnarray}
The temperature and the entropy remain related to the surface gravity and area of the black hole as in equations~(\ref{oldTD}). The $q_i$ are  gauge charges, and  $J_i$ are  angular momenta, while $\Phi_i$ and~$\Omega_i$ are their conjugate potentials and angular velocities, respectively. The black holes may have other parameters and they enter additively with their conjugates to the First Law (\ref{newTD}) in the usual way. This formalism works in multiple dimensions. Interestingly, for the static black holes, the thermodynamic volume $V$ is just the naive ``geometric'' volume of the black holes: the volume of the ball of radius $r_{+}$ (our notation for the  horizon radius in this paper)\footnote{This coincides with the definition of the volume of a static black hole proposed in ref. \cite{Parikh:2005qs}.}.

\subsection{Heat Engines and Efficiency}
\label{HHE}

In this {\it extended} black hole thermodynamics, since  the pressure and volume are now in play, alongside temperature and entropy (\ref{newTD}), it is natural to study  devices which can extract useful mechanical work from heat energy, {\it i.e.},  traditional heat engines\cite{Johnson:2014yja}. These devices were named  ``holographic heat engines'', since for  negative cosmological constant ({\it i.e.} with positive pressure, since $p=-\Lambda/8\pi$) such cycles  represent a journey through a family of holographically dual\cite{Maldacena:1997re,Witten:1998qj,Gubser:1998bc,Witten:1998zw,Aharony:1999ti} non--gravitational  field theories (at large $N_c$) defined in one  dimension fewer. Although we have holographic applications in mind for some of this work, for this paper our focus will be on the black hole side of the story, an interesting context in its own right.

So for the purposes of the gravitational theory, the working substance of the heat engine  is a particular black hole solution of the gravity system. It supplies an equation of state through the relation between its temperature $T$ and the black hole parameters defined in the usual way (we will give examples below). The precise form of all these relations depends on the type of black hole, and the parent theory 
of gravity under discussion. One may extract mechanical work from such an engine\cite{Johnson:2014yja} {\it via} the $pdV$ term in the First Law of thermodynamics in the classic way: Define a closed cycle in state space during which there is a net input heat flow $Q_{H}$, a 
 \begin{wrapfigure}{l}{0.4\textwidth}
{\centering
\includegraphics[width=1.6in]{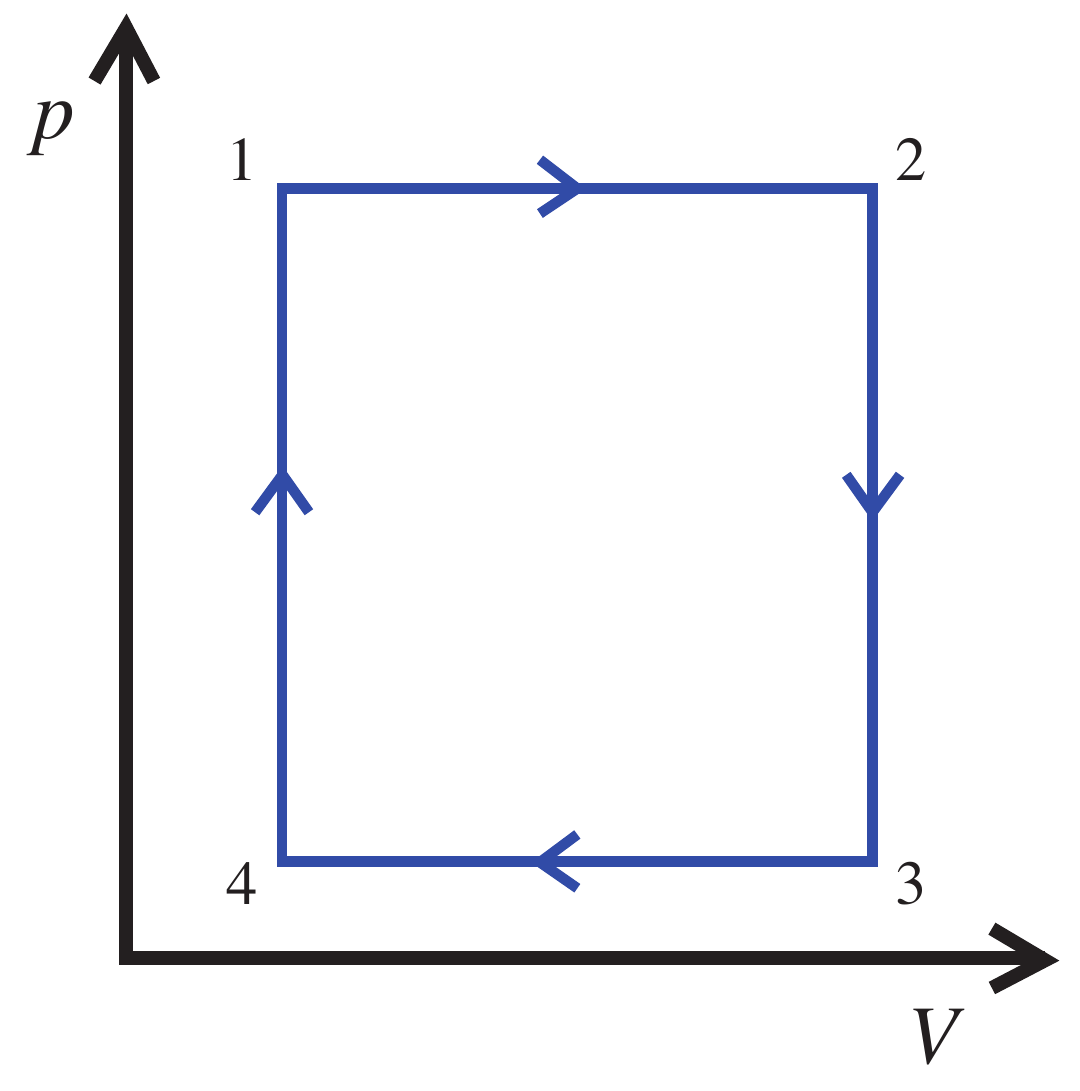} 
   \caption{\footnotesize  A prototype engine.}   \label{fig:prototype}
}
\end{wrapfigure}
net output heat flow $Q_{C}$, and a net output work $W$.  So $Q_{H}=W+Q_{C}$.  A central quantity, the efficiency for the cycle, is defined as $\eta=W/Q_{H}=1-Q_{C}/Q_{H}$. Its value is  sensitive to  the details of the equation of state of the system and also to the choice of cycle in  state space. Consider the cycle given in figure~\ref{fig:prototype}. In refs.\cite{Johnson:2014yja,Johnson:2015ekr,Johnson:2015fva} it is explained why this is a natural choice for static black holes\footnote{Refs. \cite{Belhaj:2015hha,Sadeghi:2015ksa,Caceres:2015vsa,Setare:2015yra,Zhang:2016wek,Bhamidipati:2016gel,Wei:2016hkm,Sadeghi:2016xal} have presented  further studies of such heat engines.}. For such holes, the entropy and the volume  are not independent, being both simple functions of $r_+$, the horizon radius. So isochores are adiabats,  and  so the only heat flows are along the top and bottom lines. Computing the efficiency boils down to evaluating $\int C_p dT$ along those isobars, where~$C_p$ is the specific heat at constant pressure. In general, calculation of efficiency is a difficult task to perform exactly using this approach, and  high temperature or high pressure computations are used to  get approximate results\cite{Johnson:2015ekr,Johnson:2015fva}.

Recently, however, ref. \cite{Johnson:2016pfa}  showed a much simpler way to evaluate the efficiency. The First Law is:
\begin{equation}
dH=TdS+Vdp\ , 
\end{equation}
and along the isobars, $dp=0$. Therefore the total heat flow along an isobar is simply the enthalpy change.  Normally, that might not be a useful rewriting, but in  extended gravitational thermodynamics, a precise expression for the enthalpy is readily available since it is just the   black hole mass~$M$. This results in a remarkably simple exact formula:
\begin{eqnarray} \label{SimpleMass}
\eta=1-\frac{M_{3}-M_{4}}{M_{2}-M_{1}}\ ,
\end{eqnarray}
where  the black hole mass is evaluated at each corner of the rectangle, with the labelling given in figure~\ref{fig:prototype}. $M$ is usually written as a function of $r_{+}$ and $p$. In the examples of this paper, since $V$ is a simple function of $r_{+}$, we will be easily able to write down $M$ as a function of $p$ and $V$. 

It was also shown in ref.\cite{Johnson:2016pfa} that the result (\ref{SimpleMass}) can be used as the basis for an algorithm for computing the efficiency of a cycle of {\it arbitrary} shape to any desired accuracy. Any closed shape on the state space can be approximated by tiling with a regular lattice of rectangles.
 \begin{wrapfigure}{r}{0.4\textwidth}
{\centering
\includegraphics[width=2.8in]{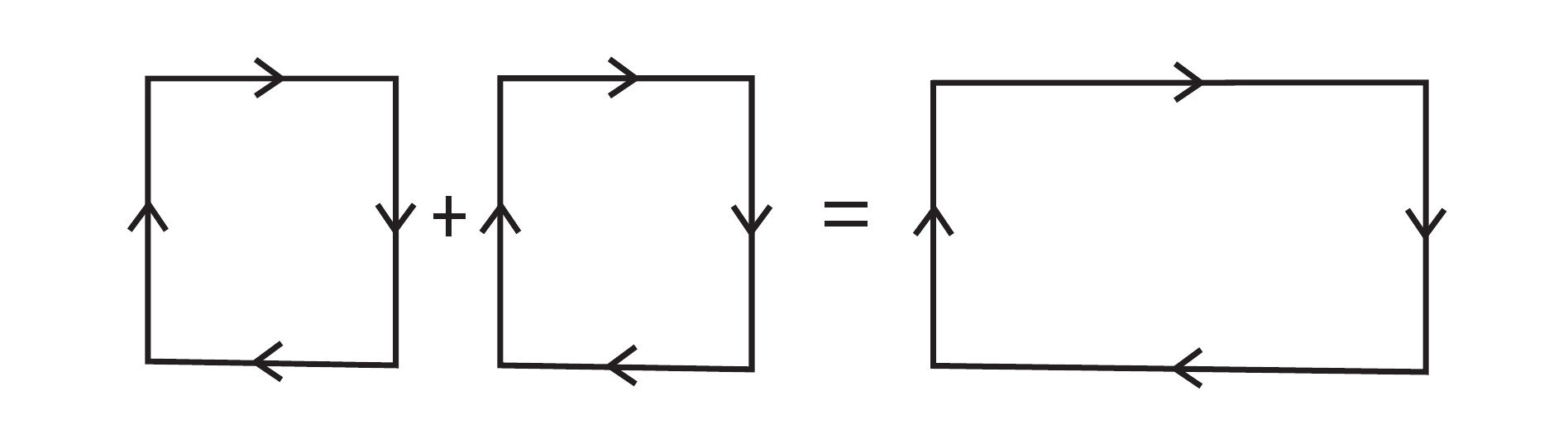} 
   \caption{\footnotesize  Adding cycles that share an edge.}   \label{fig:AddLaw}
}
\end{wrapfigure}
This is possible because cycles are additive (see figure~\ref{fig:AddLaw}). Consequently, only the cells at the edge contribute. Any mismatch between the edge of the cycle's contour and the tiling's edge can be reduced by simply shrinking the size of the unit cell. Edge cells are called hot cells if they have their upper edges open, and cold cells if they have their lower edges open.  Summing all the hot cell mass differences (evaluated at the top edges) will give $Q_{H}$ and summing all the cold cell mass differences (evaluated at the bottom edges) will yield $Q_{C}$. So the efficiency is :
\begin{eqnarray} \label{FullMass}
\eta = 1-\frac{Q_{C}}{Q_{H}}, \ \ Q_{H}=\sum_{i{\rm th} \ {\rm hot} \ {\rm cell}} (M^{(i)}_{2}-M^{(i)}_{1}), \ \ Q_{C}=\sum_{i{\rm th} \ {\rm cold} \ {\rm cell}} (M^{(i)}_{3}-M^{(i)}_{4})\ .
\end{eqnarray}
where we have labelled all cells' corners in the same way as the prototype cycle in figure~\ref{fig:prototype}.
An example with a triangular cycle was given in \cite{Johnson:2016pfa} to show the algorithm in action supporting the previous argument.

\subsection{Benchmarking}
As already stated, a given black hole, thought of as a working substance for a heat engine, supplies a particular equation of state. The efficiency will depend upon this choice. Moreover, the efficiency will also depend upon the details of the choice of cycle. For maximizing $\eta$, certain choices of cycle will be better adapted to a particular working substance (choice of black hole) than others. (For example, for the same cycle of figure~\ref{fig:prototype},  a non-static black hole will  generically have a larger $Q_H$ due to non--zero heat flows along the isochores, and therefore a smaller $\eta$.)

So a natural question arises: How does one compare the efficiency of different working substances? We have in mind a comparison that depends as little as possible on special choices of cycle. In other words,  in comparing working substances for making a heat engine, we should {\it not} choose a special cycle that favours one black hole's particular properties over another. Notice that this requirement requires us to make a choice that is in opposition to what is normally done: Cycles are usually chosen in a way that is naturally adapted to the equation of state in order to simplify computation. So we are asking that a more difficult choice of cycle be made, by necessity. 

This is where the exact formula and algorithm reviewed above come in.  We can pick a benchmark cycle of whatever shape seems appropriate and implement the algorithm to  compute $\eta$ to any desired accuracy. 

This freedom  allows us to make the following choice of benchmark: We choose the cycle to be a {\it circle} in the $p{-}V$ plane. The logic of this choice is that the circle is a simply parametrised  shape which is also unlikely to favour  {\it any} species of black hole (working substance) whatsoever. No thermodynamic variable is unchanged on any segment of the cycle, so it is, in some sense, a difficult cycle for all black holes. All that needs be specified is the origin of the circle and its radius. These properties make  it an excellent choice of benchmark.

The outline of this paper is as follows. In section \ref{sec:circle}, we set up the  circular cycle as our benchmarking tool, and explain our implementation of the exact formula and  algorithm of ref.\cite{Johnson:2016pfa}  for  calculating the efficiency.  We then discuss, in section~\ref{sec:Ideal}, a very special case of working substance: an ``ideal gas"--like system. It allows us to derive some exact results that help test our implementation, and which also set a new benchmark standard for later use. In section \ref{sec:comparison}, we  compare three examples of black holes as working substances for heat engines: Charged (Reissner--Nordstrom--like) black holes, Gauss--Bonnet black holes, and Born--Infeld  black holes. We conclude in section~\ref{sec:conclusions} with a brief discussion of future applications of our benchmarking procedure.

\section{Setting up the Circular Cycle}
\label{sec:circle}

For our circle, we implemented the algorithm and exact formula of ref.\cite{Johnson:2016pfa}, with the aid of a computer,  as follows:
Imagine that we have chosen the origin and radius, $L$, of the circle in the $p{-}V$ plane.  We next overlaid it onto the  $N\times N$ regular lattice of squares of total side length $2L$. For simplicity, we used even $N$ so that there are same number of squares both in the upper half and in the lower half of the circle. Next we computed the pressure and volume at each corner of all the squares. Using simple geometry, we determined which squares intersect the circle. We checked for cases where two squares share a common isobar and both intersect the circle. Then, if we are in the upper part of the circle, we remove the one below and keep only the upper square. We did this check in the lower half of the circle in a similar fashion. This allowed us to identify all the  hot cells and cold cells of the approximation, and their $(p,V)$ coordinates. The black hole mass is a function of pressure $p$ and volume $V$ only (with some parameters that we have already fixed), so we can compute its value at each corner. Then we use the formula (\ref{FullMass}) to give us the approximate~$\eta$ for that level of granularity. Increasing the value of $N$ makes the size of the unit cell smaller, making the path traced by the hot and cold cells a better fit to our circle, reducing the error in $\eta$. Indeed, we found that just as for the triangle prototype of ref.\cite{Johnson:2016pfa}, the efficiency converges nicely for large~$N$. (See the examples in section~\ref{sec:comparison}.)

\begin{figure}[h]
\begin{center}
{\centering
\includegraphics[width=3.5in]{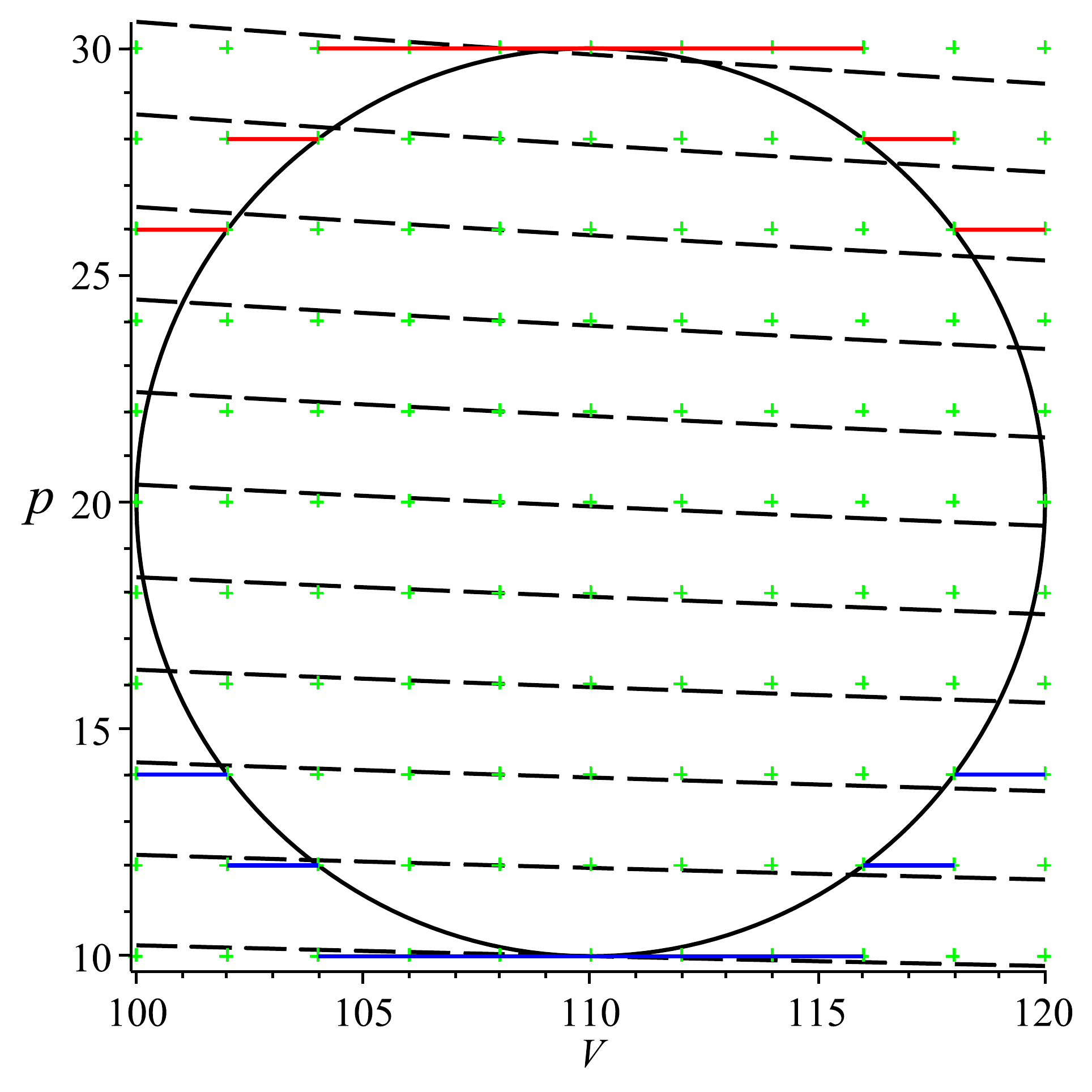} 
   \caption{\footnotesize  Example of tessellating the circular cycle for $N=10$ (100 squares). Red lines are the tops of hot cells and blue lines  are the bottoms of cold cells. As $N$ increases, these lines converge to the boundary of the circle. The dashed black lines are sample  isotherms.}   \label{fig:CCN10}
}
   \end{center}
\end{figure}

We can even do more. Since temperature is also a function of $p$ and $V$, we can compute it at each corner. Then while we run over all the cells to compute $\eta$, we can keep track of the maximum and the minimum temperatures  ($T_H$ and $T_C$) achieved in the entire cycle. Hence, we can compute the Carnot efficiency $\eta_{\rm C}=1-T_C/T_H$ for this engine. This will be a  check of our results because no  cycle can have a greater efficiency than a Carnot cycle.

Figure~\ref{fig:CCN10} shows an example for $N=10$. The green crosses show the points of the square lattice. The circle is our circular cycle. Red segments are the tops of the hot cells and blue segments are the bottoms of cold cells. The black dashed lines show a few sample isotherms determined from the underlying equation of state of the system in question. (This example is a snapshot of the Einstein--Hilbert--Maxwell case more fully explored in section~\ref{EHM}).

In choosing our benchmark cycle to compare different black holes, we should fix the circle origin and radius $L$.  Generically, the choices don't matter, as long as they are the same across the comparison. We chose $(p=20, V=110)$ here, and in the following sections, purely arbitrarily, except for making sure that we avoided any regions where the equations of state of the black holes under comparison had any multi--valuedness that would signal non--trivial phase transitions\cite{Chamblin:1999tk,Chamblin:1999hg,Kubiznak:2012wp}. Such regimes require a separate, more careful study in this heat engine context that are beyond the scope of this paper.

One might worry that since the circular cycle is presumably not even close to a cycle for which one has an  analytically computable result, if there was an error, it might not be noticed. The Carnot test  above is useful, but it is a rather weak upper bound on the efficiency. We derive some complementary tests, and a stronger (exact) bound, in the next section.

\section{The Ideal Gas Case}
\label{sec:Ideal}
 Before we proceed to study some black hole examples we briefly pause to study a simple but instructive case. It is in fact a limiting form of all of the black hole solutions we'll discuss shortly. As discussed in ref.\cite{Johnson:2015fva} it  deserves to be called an ideal gas case, and as such, sets an additional standard by which we might assess other working substances. In dimension $D$, the leading large horizon radius ($r_+$) limit of all the asymptotically anti--de Sitter black holes we will discuss is rather simple, with dependence for the mass and temperature as follows:
 \begin{equation}
\label{eq:Ideal}
M=\omega_{D-2}  \frac{r_+^{D-1}}{D-1} p+\cdots\ ,\qquad T=\frac{4r_+}{D-2}  p\cdots \ ,
\end{equation}
where $\omega_{D-2}$ is the volume ({\it i.e.,} surface area) of the unit round $S^{D-2}$ sphere.
The exact thermodynamic volume for all of the static black holes under study is:
\begin{equation}
V=\frac{\omega_{D-2}}{(D-1)} r_+^{D-1}\ ,
\label{eq:volume}
\end{equation} and so we have the familiar ``ideal gas" behaviour in this large $r_+$ limit:
\begin{equation}
p V^{1/(D-1)}\sim T\ ,
\label{eq:idealgaslaw}
\end{equation}
 a family of hyperbolae in the $p{-}v$ plane where $v=V^{1/(D-1)}$.
 This ideal gas can be obtained  as a limit for any of our black holes (in later sections) as either a large $r_+$ limit or as a  high temperature limit.  Before moving on to those cases, we can study this in its own right, taking the above as the equation of state everywhere in the $p{-}V$ plane. 
 
 Notice first that the efficiency of any cell such as the prototype of figure~\ref{fig:prototype} simplifies nicely in this case. This is because the mass is simply $M=pV$, and hence $p$ factors out in each mass difference, leaving only a volume difference. So $\eta$ for figure~\ref{fig:prototype} is just\cite{Johnson:2014yja} $\eta=1-p_4/p_1$.
 
 Turning to  the efficiency of the circle, the factorization into  sums of  volume differences means that there is no dependence of the result on the volume coordinate of the circle's origin: Any shift in the origin will cancel out everywhere. We can say even more in this case however. In fact, the terms in the sums in the algorithm~(\ref{FullMass}) are actually entirely geometrical in interpretation! For example, for a hot cell a term is of the form $p(V_2-V_1)$. This is simply the area of the rectangular strip that starts on the $V$ axis and is bounded above by the top of the cell. This is a clue to writing an {\it exact} formula for the efficiency in the case of our ideal gas. The simplest way to do it is to rewrite $\eta$ as the ratio of work to heat flowing in, $W/Q_H$. Now $W=\pi L^2$ while from our observation above, $Q_H$ is, in the large $N$ limit, exactly the area underneath the upper semi--circle of the circular path: $Q_H=\pi L^2/2+2Lp$, so our result is:
 \begin{equation}
 \eta=\frac{2\pi}{\pi+4p/L}\ , \qquad ({\rm ideal \,\,gas})
 \label{eq:idealgasefficiency}
 \end{equation}
 where $p$ is the pressure at the centre of the circle.

\begin{figure}[h]
\begin{center}
{\centering
\includegraphics[width=2.5in]{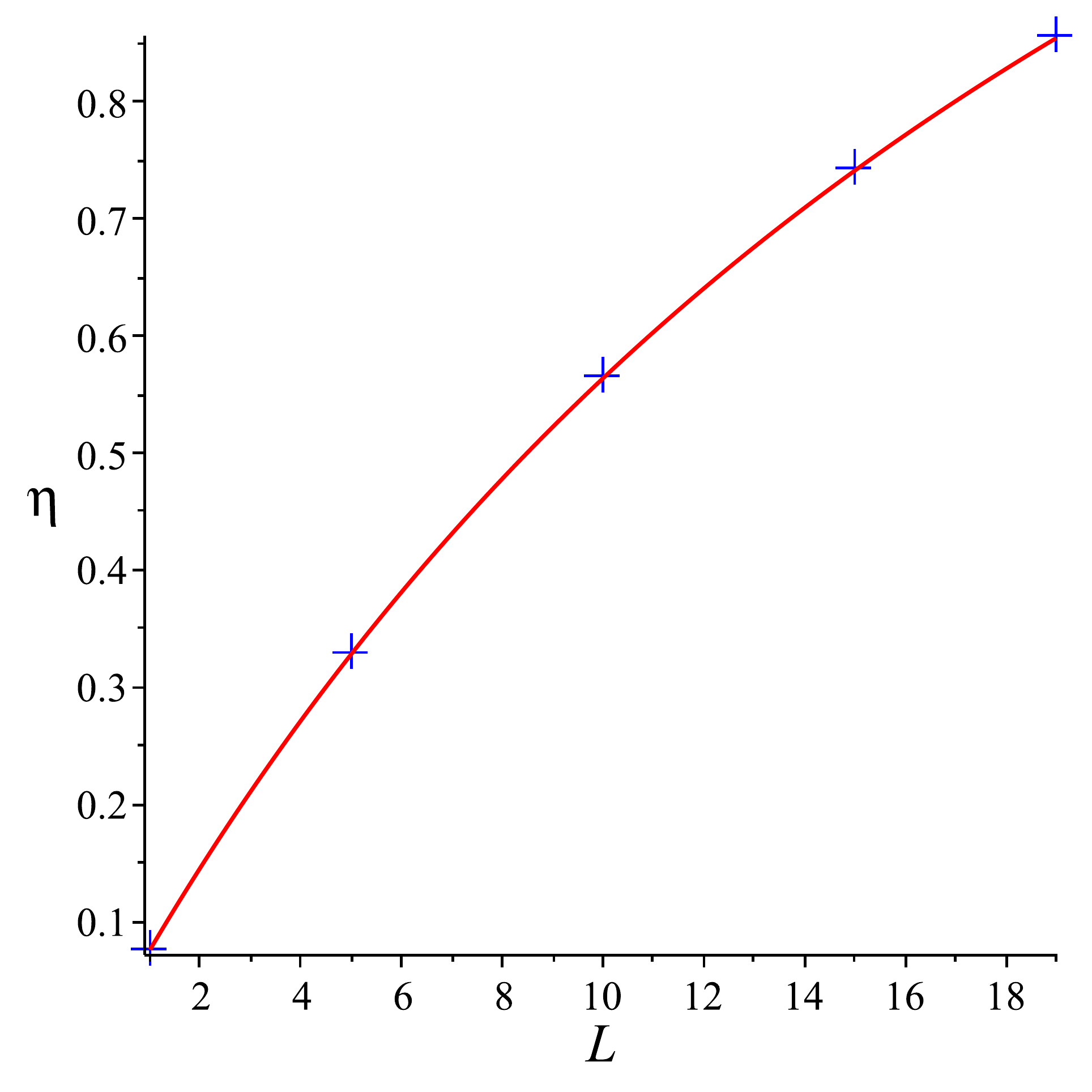} 
   \caption{\footnotesize  Efficiency of the ideal gas used in the benchmark cycle, computed as a function of circle radius $L$. The ideal gas (see text) equation of state was used in the algorithm for $N=500$ with the circle origin at $(110,20)$, and radius $L=10$. The blue  crosses  plot the result for $\eta$. The red curve is a plot of the exact result from equation~(\ref{eq:idealgasefficiency}).}   \label{fig:EffRad}
}
   \end{center}
\end{figure}

This exact formula  is rather surprising. Notably, in addition to being  independent of  $V$ it is  also independent of spacetime dimension, but the real surprise is that the algorithm assembled itself into a purely geometric result that yielded an exact formula for what is, on the face of it, a difficult shape of cycle. In fact, this exact geometrical result will work  for {\it any cycle shape}. Perhaps there can be other surprises of this sort for other systems besides this special ideal gas case.  The formula is also a rather useful check on our methods for a number of reasons. The first is that the $p$ and $L$ dependence are non--trivial predictions, and so we were obliged  to check to see if our discrete algorithm reproduces such dependence, and indeed it did.   For example, figure~\ref{fig:EffRad} shows, for $N=500$, some example points  computed by inserting the ideal gas into our algorithm. The red curve is the exact result of equation~(\ref{eq:idealgasefficiency}).

The second reason this is a strong check is that it presents a lower upper bound on our results than the upper bound given by Carnot (discussed in the previous section). Our black holes, in the regions where we study them, can be thought of as perturbations of this ideal gas case, and so we should expect that the efficiencies we obtain approach (but do not exceed) the ideal gas result.  We have, for the comparisons to come, the circle's origin at $p=20$, $V=110$, and its radius as $L=10$, for which the ideal gas efficiency is  (to six significant figures) $\eta = 2\pi/(\pi+8)\simeq 0.56394$. It is worth noting that using the discretisation algorithm to compute the ideal gas case gives $\eta \simeq 0.56588$ at $N=500$ and $\eta \simeq 0.56493$  at $N=1000$. (Moving significantly beyond $N=1000$ to see further convergence proved beyond the numerical capabilities of the system we were using.)

\section{Comparing different heat engines}
\label{sec:comparison}

We now apply our  benchmark cycle to a sampling of  different black holes acting as  working substances. We will only briefly introduce the black holes since they are well known in the literature. They were used in heat engines in refs.\cite{Johnson:2014yja,Johnson:2015ekr,Johnson:2015fva}, with some analysis and  comparison  presented there, but now we have a clearer, more systematic benchmarking procedure.

We will work in $D=5$ for definiteness (it is trivial to insert the formulae for other dimensions into our algorithm; we saw no compelling reason to present the results for other dimensions here), and our benchmark circle will be centred at  $p=20$, $V=110$, with  radius $L=10$. In each case we list  the bulk action in $D=5$ dimensions and the mass and temperature of the black hole. For static black holes  the volume $V$ is simply : $V={\pi^2}r^{4}_{+}/2$.  Also, recall that the cosmological constant~$\Lambda$ is related to pressure $p$ {\it via}  $p=-\Lambda/8\pi$, and in $D$ dimensions $\Lambda$ sets a length scale~$l$ through~$\Lambda=-(D-1)(D-2)/(2 l^2)$. So in $D=5$ dimensions, $p=3/(4 \pi  l^2)$. The mass and temperature formulae we present will have had  $l$ eliminated in favour of $p$. 

Note that in presenting our results for the efficiency, the engine's actual efficiency will be denoted by $\eta$ (without a subscript; the surrounding text will make it clear which case is being discussed) and the associated Carnot efficiency will be denoted $\eta_{\rm C}$ (again with context making it clear as to which case is being discussed). This will help us avoid a proliferation of subscripts.

\subsection{Einstein--Hilbert--Maxwell} 
\label{EHM}

The bulk action for the Einstein--Hilbert--Maxwell system in $D=5$  is\footnote{We're using the conventions of ref. \cite{Chamblin:1999tk}.}:
\begin{eqnarray} \label{EHMaction}
I=\frac{1}{16\pi}\int d^5 x\sqrt{-g}\Big(R-2\Lambda -F^2\Big) \ .
\end{eqnarray}
\begin{figure}[h]
\begin{center}
{\centering
\includegraphics[width=3.0in]{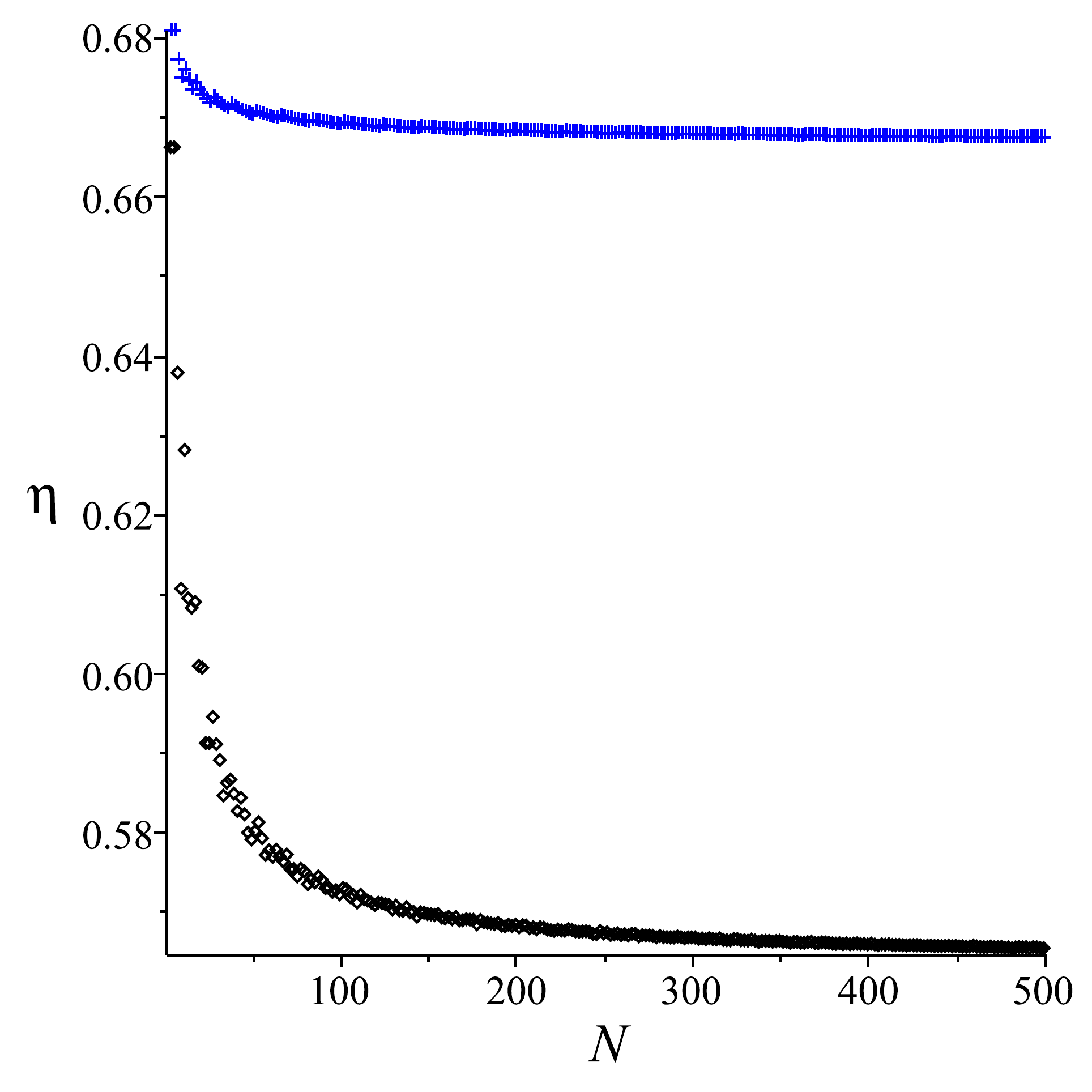} 
   \caption{\footnotesize  The efficiency of our benchmarking cycle as a function of grid size, $N$. Here Einstein--Hilbert--Maxwell black holes are used as the working substance. Blue crosses represent the  Carnot efficiency $\eta_{\rm C}$, while black squares represent $\eta$. For $N=500$, $\eta_{\rm C}$ and $\eta$ converge to 0.6674942748 and 0.5653677678 respectively.}   \label{fig:EMN500}
}
   \end{center}
\end{figure}
We can now write the mass and the temperature of the Einstein--Hilbert--Maxwell ({\it i.e., } Reissner--Nordstrom--like) black hole solution, parametrized by a charge $q$ (which we will later choose  as~$q=~0.1$):
\begin{eqnarray}
M=\frac{3\pi}{8}\Big(r^2_{+}+\frac{q^2}{r^2_{+}}+\frac{4\pi p}{3}r^4_{+}\Big)\ ,\quad {\rm and} \quad
T=\frac{1}{4\pi}\Big(\frac{16\pi p}{3}r_{+}+\frac{2}{r_{+}}-\frac{2 q^2}{r^5_{+}}\Big)\ , 
\end{eqnarray} and we can write them entirely in terms of $p$ and $V$, using  $r_{+}^4=2V/\pi^2$. Figure~\ref{fig:EMN500} shows the results of the algorithm for computing  $\eta_{\rm C}$ and $\eta$  for the benchmark circle in this case.

\subsection{Gauss--Bonnet} 
\label{GB}

In the presence of a Gauss--Bonnet sector, the action becomes\footnote{We are using the conventions of ref. \cite{Cai:2013qga}, with a slight modification of the Maxwell sector.}:
\begin{eqnarray} \label{GBaction}
I=\frac{1}{16\pi}\int d^5 x\sqrt{-g}\Big(R-2\Lambda +\alpha_{GB} (R_{\gamma \delta \mu \nu }R^{\gamma \delta \mu \nu } -4R_{\mu \nu }R^{\mu \nu } +R^2)-F^2\Big)
\end{eqnarray}
where $\alpha_{GB}$ is the Gauss--Bonnet parameter which has dimensions of $({\rm length})^2$. If we set $\alpha_{GB}=0$ in (\ref{GBaction}) we go back to the previous case of Einstein--Hilbert--Maxwell system (\ref{EHMaction}).

\begin{figure}[h]
\begin{center}
{\centering
\includegraphics[width=3.0in]{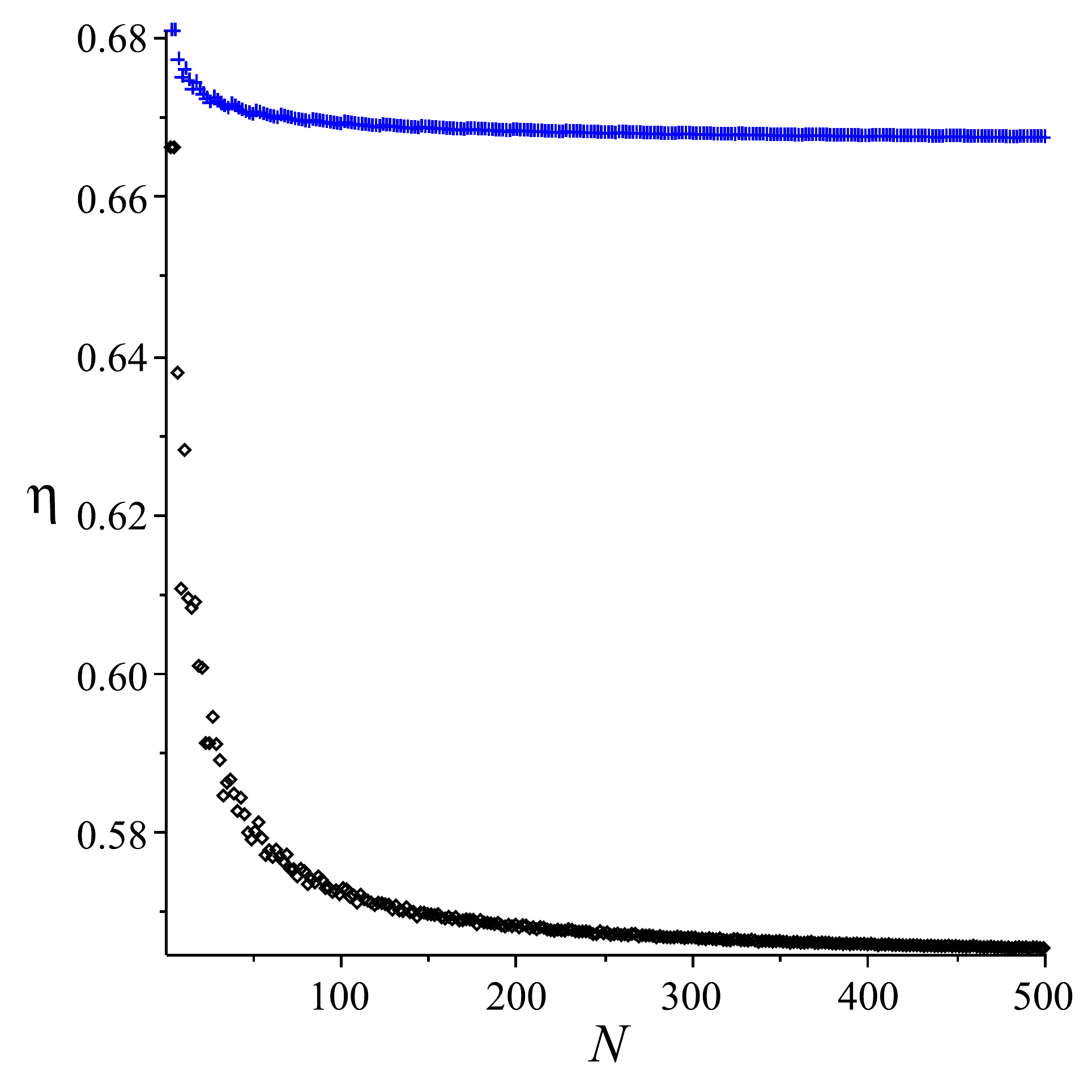} 
   \caption{\footnotesize  The efficiency of our benchmark cycle as a function of grid size, $N$. Here Gauss--Bonnet black holes are used as the working substance. Blue crosses represent the  Carnot efficiency $\eta_{\rm C}$, while black squares represent $\eta$. For $N=500$, $\eta_{\rm C}$ and $\eta$ converge to 0.6674954523 and 0.5653678245 respectively.}   \label{fig:GBN500}
}
   \end{center}
\end{figure}

 The mass and  temperature of the black hole, parametrized by $q$ and $\alpha$ are:
\begin{eqnarray}
M=\frac{3\pi}{8}\Big(\alpha +r^2_{+}+\frac{q^2}{r^2_{+}}+\frac{4\pi p}{3}r^4_{+}\Big)\ , \quad {\rm and} \quad
T=\frac{1}{4\pi(1+\frac{2\alpha}{r^2_{+}})}\Big(\frac{16\pi p}{3}r_{+}+\frac{2}{r_{+}}-\frac{2 q^2}{r^5_{+}}\Big)\ ,
\end{eqnarray}
where  $\alpha=2 \alpha_{GB}$. We will again work with $q=0.1$ and we choose a sample value of the coupling as $\alpha=0.001$.
See figure~\ref{fig:GBN500} for $\eta_{\rm C}$ and $\eta$ from the benchmark analysis.

\subsection{Born--Infeld} 
\label{BI}

The so--called\footnote{See {\it e.g.} the remarks in ref.\cite{Johnson:2015fva} about the terminology} Born--Infeld action \cite{Born:1933,Born:1934ji,Born:1934gh} is a non-linear generalization of the Maxwell action, controlled by the parameter $\beta$ :
\begin{eqnarray} \label{BIsector}
{ \cal L}(F)=4 \beta^2 \Big(1-\sqrt{1+\frac{F^{\mu \nu}F_{\mu \nu}}{2 \beta^2}}\Big)
\end{eqnarray}
If we take the limit $\beta \rightarrow \infty$ in (\ref{BIsector}) we recover old Maxwell action. The Einstein--Hilbert--Born--Infeld bulk action in $D=5$ is obtained by replacing the Maxwell sector in equation~\ref{EHMaction} with this action.
\begin{figure}[h]
\begin{center}
{\centering
\includegraphics[width=3.0in]{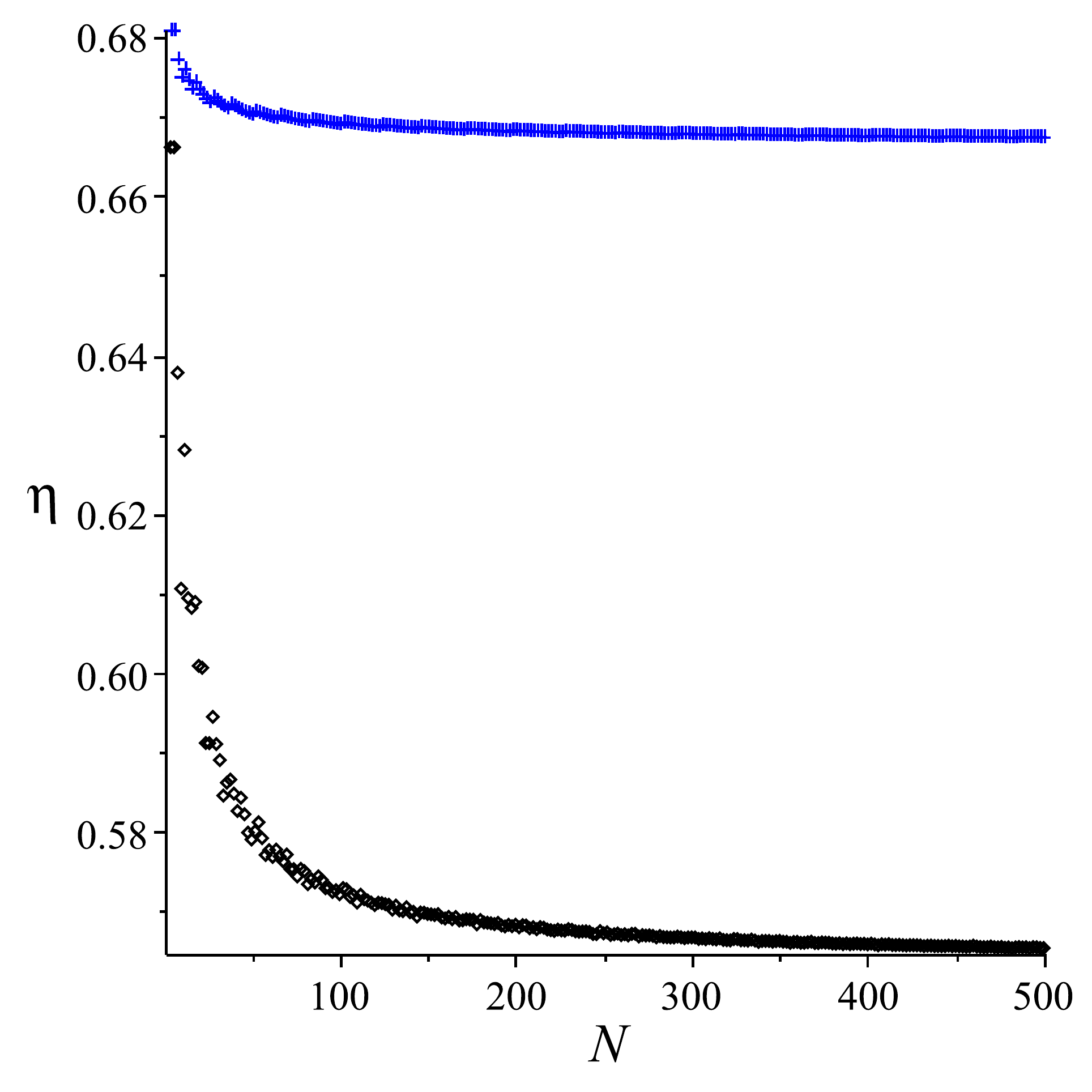} 
   \caption{\footnotesize  The efficiency of our benchmark cycle as a function of grid size, $N$. Here Born--Infeld black holes are used as the working substance. Blue crosses represent the  Carnot efficiency $\eta_{\rm C}$, while black squares represent $\eta$. For $N=500$, $\eta_{\rm C}$ and $\eta$ converge to 0.6674942730 and 0.5653678967 respectively.}   \label{fig:BIN500}
}
   \end{center}
\end{figure}
The exact results for the Born--Infeld black hole's mass and temperature are known\footnote{See refs. \cite{Fernando:2003tz,Cai:2004eh,Dey:2004yt} for further details.}, but for our purposes, it is enough to expand them in $1/\beta$, keeping only  leading non--trivial terms. For the mass:
\begin{eqnarray}
M=\frac{3 \pi}{8}\Big(r^2_{+}+\frac{4 \pi p}{3}r^4_{+}+\frac{q^2}{r^2_{+}}(1-\frac{9 q^2}{16 \beta^2 r^6_{+}})\Big)+\mathcal{O}\Big(\frac{1}{\beta^4}\Big)
\end{eqnarray}
and the temperature:
\begin{eqnarray}
T=\frac{1}{4 \pi}\Big(\frac{16 \pi p}{3}r_{+}+\frac{2}{r_{+}}-\frac{2 q^2}{r^5_{+}}(1-\frac{3 q^2}{4 \beta^2 r^6_{+}})\Big)+\mathcal{O}\Big(\frac{1}{\beta^4}\Big)
\end{eqnarray}
The exact formulae are computationally intensive, and for any significant $N$, there are far too many computations  to allow computation of the efficiency in a reasonable amount of time (especially in $D=5$) and so we chose to make this truncation at the outset. We  worked with $q=0.1$ and $\beta=0.1$ in our benchmark studies, the results of which are shown in  figure~\ref{fig:BIN500} for $\eta_{\rm C}$ and $\eta$.

\subsection{Comparison/Observations}
\label{Comp}
\begin{figure}[h]
\begin{center}
{\centering
\includegraphics[width=2.0in]{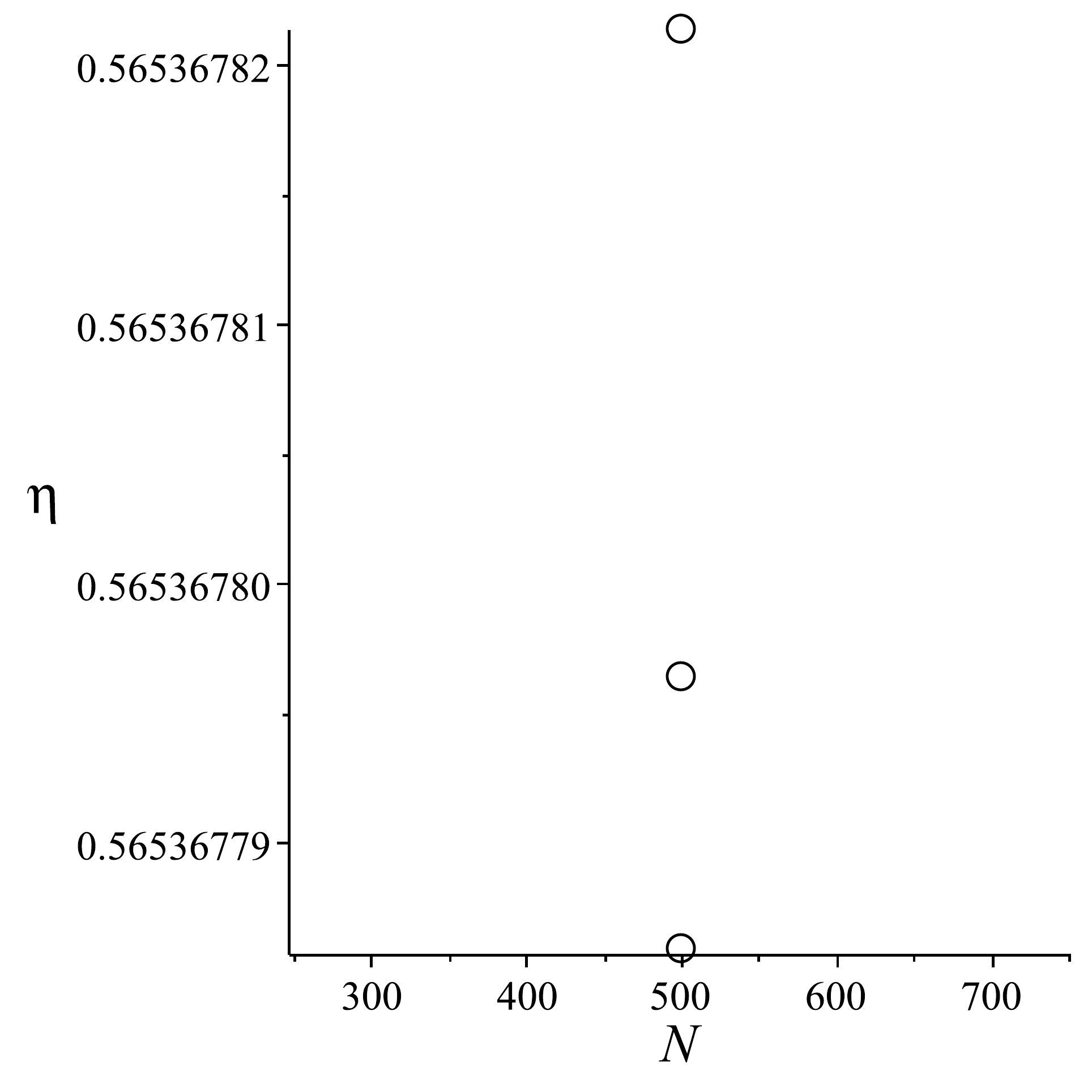} 
   \caption{\footnotesize  The efficiencies of Einstein--Hilbert-Maxwell (lowest), Gauss--Bonnet (highest) and Born--Infeld (middle) black hole heat engines for $N=500$ with circle origin at $(110,20)$ and radius $L=10$. For additional comparison, the ideal gas case of section~\ref{sec:Ideal} has $\eta\simeq 0.56588$ at $N=500$, and is $\eta=2\pi/(\pi+8)$   exactly.}   \label{fig:PC}
}
   \end{center}
\end{figure}

In figure~\ref{fig:PC} we gather all the efficiencies computed using the benchmark cycle together. The Gauss--Bonnet and Born--Infeld cases, thought of as perturbations of the Einstein--Maxwell case,  have higher efficiencies, although it is interesting that the differences begin to show only in the 8th significant figure, for the parameter values chosen for $\alpha$ and $\beta$.  We explored other parameter values (while making sure to stay in the physical range allowed by reality of the mass for the Gauss--Bonnet case) and found a very weak dependence of the efficiency as they varied. (This all matches observations made in refs.\cite{Johnson:2015ekr,Johnson:2015fva} in the high temperature limit.) They all in turn have significantly lower efficiency than the ideal gas case listed at the end  of  section~\ref{sec:Ideal}. 

%
%

\section{Discussion}
\label{sec:conclusions}

We've defined a new way of comparing different black holes, given meaning in the context of defining black hole heat engines\cite{Johnson:2014yja} in extended thermodynamics.  Our benchmarking allowed us to compare four important cases against each other, and we found results consistent with earlier studies reported in refs.\cite{Johnson:2015ekr,Johnson:2015fva}, but here we've established a more robust  framework for comparison (a standard circular cycle) facilitated by the exact formula and algorithm of ref.\cite{Johnson:2016pfa}. Along the way, we found a fascinating case where the algorithm itself collapses to another exact result, this time the exact efficiency of an ``ideal gas" example.  It would be fascinating  to see if other exact results of this kind can be obtained for other non--trivial systems. It would be  interesting to study other black holes using this same benchmarking scheme in order to compare more properties of heat engine working substances. Extend all this  to non--static cases would be particularly worthwhile. Finally, the possible applications of all of this to holographically dual strongly coupled field theories is worth exploring.   We hope to  report on some of this elsewhere.

\section*{Acknowledgments}

We would like to thank   the  US Department of Energy for support under grant DE--SC0011687. CVJ would like to thank Amelia for her patience and support.


\providecommand{\href}[2]{#2}\begingroup\raggedright\endgroup

\end{document}